# EFFECT OF PROGRAMMED DEATH LIGAND-1 EXPRESSION LEVELS ON CLINICOPATHOLOGIC FEATURES AND SURVIVAL IN SURGICALLY RESECTED SARCOMATOID LUNG CARCINOMA

Yetkin Ağaçkıran, MD, Associate Professor in Pathology

Department of Pathology, Health Sciences University Atatürk Chest Diseases and Chest Surgery, Ankara, Turkey.

Funda Aksu, MD, Associate Professor in Pulmonolgy.

Department of Chest Diseaes, Health Sciences University Atatürk Chest Diseases and Chest Surgery, Ankara, Turkey.

Nalan Akyürek, MD, Professor in Pathology Department of Pathology,

Gazi University School of Medicine, Ankara, Turkey.

Caner Ercan, MD, Consultant in Pathology.

Instute of Pathology, University Hospital Basel, Switzerland.

Mustafa Demiröz, MD, Consultant in Chest Surgery

Department of Chest Surgery, Health Sciences University Atatürk Chest Diseases and Chest Surgery, Ankara, Turkey.

Kurtuluş Aksu, MD, Associate Professor in Pulmonology, Clinical Immunology and Allergy.

Department of Chest Diseases, Division of Immunology and Allergy, Health Sciences University Atatürk Chest Diseases and Chest Surgery, Ankara, Turkey.

**Running Title:** PDL1 EXPRESSION IN SARCOMATOID LUNG CARCINOMA
**Corresponding author**: Funda Aksu, MD

Phone: 90 5326261877

Fax number: 90 3123572135

Address: Department of Chest Diseaes, Health Sciences University Atatürk Chest Diseases and Chest Surgery, Pınarbaşı Mahallesi, Sanatoryum Caddesi 06280, Keçiören, Ankara, Turkey.

email: fundayardim@gmail.com

**Acknowledgements:** PD-L1 staining of specimens was funded by Health Sciences University, Atatürk Chest Diseases and Chest Surgery Hospital.

Submitted version: 15-Sep-2019

## Abstract

**Aim:** To define PD-L1 expression rates in sarcomatoid lung carcinomas and to assess clinicopathologic features and survival rates of subjects with PD-L1 positive lung carcinomas.

**Methods:** PD-L1 expression was evaluated in 65 surgically resected sarcomatoid carcinoma cases. Clinicopathologic features of cases with PD-L1 positive and negative tumors were compared. Kaplan-Meier survival analysis were performed Multiple Cox's proportional hazard regression analysis was applied to determine independent predictors which affected overall survival.

**Results:** PD-L1 antibody positivity is encountered in 72.3% of surgically resected sarcomatoid lung carcinomas. Regarding histopathologic subtypes PD-L1 expression was positive in 80.4% of pleomorphic carcinomas, 62.5 % of spindle and/or giant cell carcinomas and 16.7% of carcinosarcomas. Pleural invasion is observed in 68.1% of PD-L1 positive cases compared to 27.8% of PD-L1 negative ones (p=0.008). The overall mean life expectancy (95%CI) is 53.6 (41.0-65.2) months in sarcomatoid lung carcinomas. Only significant factor related to overall survival is pathological stage of the tumor.

**Conclusions:** The present study reveals data on PD-L1 positivity in a large number of sarcomatoid lung carcinoma cases consisting of four histopathological subtypes; which are pleomorphic carcinoma, spindle cell carcinoma, giant cell carcinoma and carcinosarcoma. PD-L1 expression tended to be negative in carcinosarcomas, whereas PD-L1 expression was higher in pleomorphic carcinomas. PD-L1 positivity was not a significant factor related to overall survival in sarcomatoid lung carcinomas.

**Introduction**

Sarcomatoid carcinoma of lung is an extremely rare subgroup of lung neoplasm. Although it is reported to be seen 2-3% in surgical series, large epidemiologic researches report a ratio of 0.5%.[1] These tumors are highly aggressive and they tend to produce distant metastases on unusual locations. Survival rates of these tumors are also significantly lower than the other types of non-small cell lung carcinoma.[2] Also, these aggressive tumors are resistant to chemotherapy and radiation therapy. The biology of tumor progression and chemo- or radio-resistance mechanisms of these tumors is not clear.[3,4] According to WHO classification of tumors, sarcomatoid carcinoma has five subtypes; pleomorphic carcinoma, spindle cell carcinoma, giant cell carcinoma, carcinosarcoma and pulmonary blastoma.[5]

Sarcomatoid carcinoma has higher rates of PD-L1 expression then other types of NSCLC.[4,6] PD-L1 is a ligand which is expressed as a transmembrane protein on normal cells.[7] When PD-L1 binds to PD1 receptors expressed by CD8+ T lymphocytes T lymphocyte are activated.[8,9] This ligand-receptor binding between PD1 and PD-L1 ensures recognition of host cells by T lymphocytes and obstructs cytotoxicity. Similar ligand-receptor binding takes place in cases of PD-L1 expressing tumor cells resulting in suppression of T lymphocyte cytotoxicity. Once PD-L1 expression is detected on tumor cells immunohistochemically, immune checkpoint inhibitors can be used in treatment. Antibodies against PD-L1 blocking the ligand-receptor binding enhance activation of T lymphocytes and tumor cell cytotoxicity.[10-16] Immune checkpoint inhibitors are promising in treatment of PD-L1 positive sarcomatoid carcinoma as they are related to better prognosis and survival compared to conventional therapies.[12,17-19] This study aims to define PD-L1 expression rates in sarcomatoid lung carcinoma with immunohistochemistry and to assess clinicopathologic features of cases with PD-L1 positive lung carcinomas as well as survival.

**Material and Methods**

A retrospective study was designed to evaluate surgically resected sarcomatoid carcinoma cases pathologically diagnosed in Health Sciences University Atatürk Chest Diseases and Chest Surgery Training and Research Hospital and Gazi University Medical School Pathology departments in between 2011 and 2018. Of 67 sarcomatoid carcinoma cases, 65 had full clinical and demographical data in hospital files and thus were included in this study. The hematoxylin-eosin and immunohistochemically stained slides were re-evaluated for diagnosis and sub-typing, by two separate pathologists. Re-evaluation was made in the base of current World Health Organization classification of lung tumors. Immunohistochemical study was performed on 3 to 4 μm thick paraffin embedded tissue sections placed on poly-L-Lysine coated slides. These sections were kept in a 45 °C incubator for 24 hours. For detecting PD-L1 antibody (SP263, rabbit monoclonal antibody, Ventana) by immunohistochemically, standard antigen retrieval methods (deparaffinization, cell restoration, CC1-64min, pre-primer peroxidase inhibitor-primer antibody; 16min-36°C) were applied in Benchmark GX Immunohistochemistry device (Ventana) and then Optiview DAP Immunohistochemistry detection kit (Optiview HQ Universal linker

Optiview HRP Multimer,Ventana) was used. Samples were counterstained for 4 minutes with hematoxylin II and postcounterstained for 4 minutes. Placenta was used as positive control for PD-L1 staining.

PD-L1 expression was evaluated in terms of percentage of membranous or cytoplasmic type of staining. Accordingly, PD-L1 expression was evaluated as strong (>50%), moderate (25-50%), weak (<25%): or none (0%). A membranous and/or cytoplasmic staining ≥ 25% was considered as positive staining for PD-L1.[20] Pleural invasion in cases was classified according to 7th edition of the TNM classification for lung cancer as; PL0, tumor with no pleural involvement beyond its elastic layer; PL1, tumor that invades beyond the elastic layer of the visceral pleura but is not exposed on the pleural surface; PL2, tumor that invades to the pleural surface and PL3, tumor that invades to the parietal pleura. Pathological stage of surgically resected tumors were defined according to TNM classification of lung cancer.[21,22]

**Statistical Analysis**

Descriptive statistics were shown as number of cases and percentages. Categorical data were analyzed by continuity corrected Chi-square or Fisher's exact test, where applicable. Mann Whitney U test was applied for the comparisons of ordinal variables. Kaplan-Meier survival analysis with Log-Rank test was used to examine prognosis on overall survival. The crude survival ratios, 1- 3- and 5-year cumulative survival rates, mean expected life duration and 95% confidence intervals were also calculated. Multiple Cox's proportional hazard regression analysis was applied for determining the best independent predictor(s) which mostly affected on overall survival after adjustment for clinically important factors. Any variable whose univariable test had a p value<0.25 was accepted as a candidate for the multivariable model. Relative risks and 95% confidence intervals for each independent variable were also calculated. Data analysis was performed by using IBM SPSS Statistics version 17.0 software (IBM Corporation, Armonk, NY, USA). A p value less than 0.05 was considered as statistically significant.

**Results**

This study involved 65 subjects with age; median (min-max): 61 (36-80) years and gender distribution male/female: 60 (92.3%)/5 (7.7%). The study population revealed ≥25% staining of tumor cells with PD-L1 antibody staining in 47 (72.3%) cases (PD-L1 positive group). Among these PD-L1 positive specimens 46 (97.9%) had PD-L1 antibody staining at higher intensity (≥50%). Weak staining (1-25%) and no staining with PD-L1 antibody was detected in 11(16.9%) and 7 (10.8%) specimens respectively (PD-L1 negative group). Positive staining with PD-L1 antibody was detected in 80.4% of pleomorphic carcinomas, 62.5% of spindle and/or giant cell carcinomas and 16.7% of carcinosarcoma (Table 1). Immunohistochemical samples of PD-L1 expression in sarcomatoid carcinomas are demonstrated in Figure 1.

Demographical and clinicopathologic characteristics of PD-L1 negative and positive subjects are demonstrated in Table 2. Pleural invasion was observed at higher rates in PD-L1 positive cases compared to PD-L1 negative cases. Regarding histopathologic subtype of surgically resected sarcomatoid lung carcinomas, PD-L1 expression tended to be negative in carcinosarcomas, whereas PD-L1 expression was higher in pleomorphic carcinomas. Age distribution, gender, smoking status, tumor size, lymph node metastasis and pathological stage were not statistically different between PD-L1 negative and positive cases.

Kaplan-Meier analysis with log-rank test was used to determine survival rates and to identify risk factors related to poor overall survival. The analysis revealed that overall mean life expectancy (95%CI) is 53.6 (41.0-65.2) months and only significant factor related to overall survival is pathological stage of the tumor (p=0.004). Stage III tumors had worse prognosis compared to stage I and II tumors (p=0.015 and 0.004 respectively). Although PD-L1 positive cases had worse prognosis compared to PDL1 negative cases no statistically significant difference was found between the overall survivals of the two groups (Table 3). Kaplan Meier overall survival curves of PD-L1 negative and positive sarcomatoid lung carcinoma cases are demonstrated in Figure 2. Factors found to be related to overall survival according to univariate analysis at statistical significance of. 0.025 value and factors that might clinically be related to overall survival were evaluated in Cox regression analysis. The analysis results revealed only independent factor related to prognosis is pathological stage (Table 4)

**Discussion**

Sarcomatoid carcinomas are extremely rare lung tumors whose survival rates and prognosis are significantly worse than the other types of non-small cell carcinoma. Literature data reveal gaining knowledge on PD1/PD-L1 binding play key role from escaping the immune response in these tumors leading to inhibition of T lymphocyte cytotoxicity. Accordingly, PD-L1 expression is critical in these aggressive tumors since therapies targeting PD1/PD-L1 pathway are emerging treatment of sarcomatoid carcinomas .[15]

According to the present study PD-L1 positivity is encountered in 72.3% of sarcomatoid carcinoma cases. The rate was even higher in pleomorphic carcinoma subgroup as 80.4%, relatively lower in spindle cell carcinoma subgroup as 62.5% and extremely lower in carcinosarcoma subgroup as 16.7%. No difference in demographical features of PD-L1 positive and negative cases were observed regarding age, gender and smoking status. When clinicopathologic properties of PD-L1 positive and negative groups are compared, pleural invasion was only statistically significant parameter. PDL1 expression is higher in tumors with pleural invasion (PL1,2 and 3) compared to no pleural invasion.

PD-L1 levels in sarcomatoid lung carcinoma were first reported in 2013 by Velcheti et al. They have investigated PD-L1 levels in lung carcinoma cases of which 13 were sarcomatoid carcinoma and reported the rate of PD-L1 expression in sarcomatoid carcinoma as 69, 2%.[4] Kim et al studied PD-L1 expression in 41 pleomorphic lung carcinoma cases and reported positive staining at a rate of 90, 2% .[23] Chang et al also reported high rates of PD-L1 positivity up to 70,5 % in 122 pleomorphic carcinoma cases.[24] Vieira et al report about 75 sarcomatoid carcinoma including 59 pleomorphic carcinoma cases showed that the rate of PD-L1 expression is 53, 5% .[25] Yvorel et al reported PD-L1 expression rate as 75%, in 36 sarcomatoid carcinoma .[26]

Imanishi et al were reported an expression rate of PD-L1 as 69.2-76, 9% in pleomorphic carcinoma cases .[27] Clones used as PD-L1 antibody were E1L3N in three, 5H1 in two and proteintech in one of these reports above.

The present study assessed PD-L1 positivity in sarcomatoid carcinomas with use of SP263 clone for immunohistochemical staining. PD-L1 clone SP263 is a rabbit monoclonal primary antibody used to assess PD-L1 expression. It has recently received CE-mark label to evaluate treatment decisions in subjects with non-small cell carcinoma patients being considered for immune check point inhibitors according to comparative studies with other currently available PD-L1 assays.[28] In spite of different cut-off values even for same clone of PD-L1 antibodies in the previous studies in the present study a cut-off value for PD-L1 positivity was determined as ≥ 25% staining with SP263 clone.

The remarkable clinicopathological finding of the study is the higher PD-L1 positivity in tumors with pleural invasion compared to non-invasive tumors. Association with PDL1 expression and pleural invasion has also been recently reported by Naito et al and they concluded that PD-L1 expression might be considered as a poor prognostic factor.[29] In their study Naito et al studied 35 cases of surgically resected pleomorphic carcinoma. The present work has studied not only pleomorphic carcinomas but sarcomatoid carcinomas including spindle cell carcinoma and/or giant cell carcinoma and carcinosarcoma subgroups also. Association between PD-L1 positivity and pleural invasion was assessed in a study group with a larger population.

Data regarding PD-L1 positivity in lung carcinomas and survival is conflicting. Except one study reporting better prognosis in PD-L1 positive lung carcinomas, PD-L1 expressing carcinomas are related to poor prognosis .[23-27] In the present study mean life expectancy was shorter in PD-L1 positive cases compared to PD-L1 negative cases. However, there was no statistically significant difference between PD-L1 positive and PD-L1 negative groups in terms of overall survival. Overall survival analysis evaluated by Cox regression analysis revealed that the only independent factor related to prognosis was pathological stage

The present study is an important report on clinicopathologic features of surgically resected PD-L1 positive sarcomatoid lung carcinomas. It includes high number of sarcomatoid lung carcinoma cases not only restricted to pleomorphic carcinomas but also other histopathological subtypes. With this study PD-L1 positivity rates are studies in 4 histopathological subtypes namely; pleomorphic carcinoma, spindle cell carcinoma, giant cell carcinoma and carcinosarcoma. PDL-1 positivity is studied with PD-L1/clone SP263 antibody for the first time in sarcomatoid lung carcinomas.

**Acknowledgements:** PD-L1 staining of specimens was funded by Health Sciences University, Atatürk Chest Diseases and Chest Surgery Hospital.

**Conflicts of Interest:** The authors have no conflicts of interest to declare.

**Ethical Statement:** The study was approved by Health Sciences University Atatürk Chest Diseases and Chest Surgery Education and Research Hospital Medical Expert Board (07.11.2017/572).

**Table 1.** PD-L1 Expression in Sarcomatoid Lung Carcinomas (n=65)

|  | PD-L1 Negative n=18 (27.7 %) | | PD-L1 Positive n=47 (72.3 %) | | |
| --- | --- | --- | --- | --- | --- |
|  | 0 % | <25 % | 25-50 % | ≥ 50 % | Positive % |
| **Pleomorphic Carcinoma (n=51)** | **4 (7.8)** | **6 (11.8)** | **1 (2.0)** | **40 (78.4)** | **41 (80.4)** |
| Adenocarcinoma with SCC and / or GCC (n=28) | 2 (7.1) | 2 (7.1) | 1 (3.6) | 23 (82.1) | 24 (85.7) |
| Squamous cell carcinoma with SCC and / or GCC (n=15) | 2 (13.3) | 4 (26.7) | 0 (0) | 9 (60.0) | 9 (60.0) |
| Large cell carcinoma with SCC and / or GCC (n=7) | 0 (0) | 0 (0) | 0 (0) | 7 (100.0) | 7 (100.0) |
| Adenosquamous carcinoma with SCC and / or GCC (n=1) | 0 (0) | 0 | 0 (0) | 1 (100.0) | 1 (100.0) |
| **Spindle and/or Giant cell carcinoma (n=8)** | **2 (25.0)** | **1 (12.5)** | **0 (0)** | **5 (62.5)** | **5 (62.5)** |
| **Carcinosarcoma (n=6)** | **1 (16.7)** | **4 (66.7)** | **0** | **1 (16.7)** | **1 (16.7)** |
| Total (n=65) | 7 (10.8) | 11 (16.9) | 1 (1.5) | 46 (70.8) | 47 (72.3) |

Data are expressed in n(%). SCC:Spindle cell carcinoma, GCC: Giant cell carcinoma

**Table 2.** Clinicopathologic features of patients with sarcomatoid lung carcinomas

| | PD-L1 expression | | | |
|---|---|---|---|---|
| | PD-L1 negative (n=18) | PD-L1 positive (n=47) | Total (n=65) | p-value |
| **Age** | | | | 0.565† |
| < 60 | 9 (50.0%) | 18 (38.3%) | 27 (41.5%) | |
| ≥ 60 | 9 (50.0%) | 29 (61.7%) | 38 (58.5%) | |
| **Gender** | | | | 0.125‡ |
| Male | 15 (83.3%) | 45 (95.7%) | 60 (92.3%) | |
| Female | 3 (16.7%) | 2 (4.3%) | 5 (7.7%) | |
| **Smoking status** | | | | 0.480‡ |
| Smoker | 14 (77.8%) | 40 (85.1%) | 54 (83.1%) | |
| Non-smoker | 4 (22.2%) | 7 (14.9%) | 11 (16.9%) | |
| **Tumour size** | | | | 0.489‡ |
| ≤ 3 | 5 (27.8%) | 8 (17.0%) | 13 (20.0%) | |
| > 3 | 13 (72.2%) | 39 (83.0%) | 52 (80.0%) | |
| **Lymph node metastasis** | | | | >0.999‡ |
| Absent | 13 (72.2%) | 35 (74.5%) | 48 (73.8%) | |
| Present | 5 (27.8%) | 12 (25.5%) | 17 (26.2%) | |
| **Pleura invasion** | | | | **0.008†** |
| PL0 | 13 (72.2%) | 15 (31.9%) | 28 (43.1%) | |
| PL1,PL2,PL3 | 5 (27.8%) | 32 (68.1%) | 37 (56.9%) | |
| **Pathological stage** | | | | 0.343¶ |
| I | 6 (33.3%) | 10 (21.3%) | 16 (24.6%) | |
| II | 7 (38.9%) | 20 (42.6%) | 27 (41.5%) | |
| III | 5 (27.8%) | 17 (36.2%) | 22 (33.8%) | |
| **Histopatologic Subtype** | | | | |
| Pleomorphic carcinoma | 10 (55.6%) | 41 (87.2%) | 51 (78.5%) | **0.015‡** |
| Spindle and/or Giant cell carcinoma | 3 (16.7%) | 5 (10.6%) | 8 (12.3%) | 0.675‡ |
| Carcinosarcoma | 5 (27.8%) | 1 (2.1%) | 6 (9.2%) | **0.005‡** |

Data are expressed as n (%). †Continuity corrected Chi-square test, ‡Fisher's exact test, ¶Mann Whitney U test

Table 3. The results of univariate survival analyses

| | Crude survival ratio (%) | Survival rates (%) | | | Mean life expectacy (95% CI) (months) | Log-Rank | p-value |
|---|---|---|---|---|---|---|---|
| | | *1-year* | *3-year* | *5-year* | | | |
| **Age** | | | | | | 0.337 | 0.562 |
| *< 60* | 57.7 | 69.2 | 54.5 | 54.5 | 69.2 54.5 (38.2-70.7) | | |
| *≥ 60* | 45.9 | 73.0 | 48.5 | 34.0 | 73.0 48.5 (35.2-64.1) | | |
| **Gender** | | | | | | 0.252 | 0.615 |
| *Male* | 51.7 | 72.4 | 51.6 | 41.3 | 51.6 (41.9-66.3) | | |
| *Female* | 40.0 | 60.0 | 40.0 | 40.0 | 27.4 (6.7-48.1) | | |
| **Smoking status** | | | | | | 0.295 | 0.587 |
| *Smoker* | 54.5 | 54.5 | 54.5 | - | 28.6 (16.1-41.2) | | |
| *Non-smoker* | 50.0 | 75.0 | 51.3 | 41.6 | 54.5 (42.0-66.9) | | |
| **Tumour size** | | | | | | 0.521 | 0.470 |
| *≤ 3* | 58.3 | 83.3 | 62.5 | 50.0 | 52.2 (33.8-70.6) | | |
| *> 3* | 49.0 | 68.6 | 48.1 | 40.1 | 51.6 (38.6-64.7) | | |
| **Lymph node metastasis** | | | | | | 0.856 | 0.355 |
| *Absent* | 56.5 | 71.7 | 57.2 | 46.8 | 54.7 (41.7-67.6) | | |
| *Present* | 35.3 | 70.6 | 38.5 | 30.8 | 45.0 (25.2-64.7) | | |
| **Pleura invasion** | | | | | | 2.528 | 0.112 |
| *PL0* | 59.3 | 85.2 | 57.1 | 50.0 | 63.0 (46.0-80.0) | | |
| *PL1,PL2,PL3* | 44.4 | 61.1 | 46.4 | 36.1 | 44.5 (30.2-58.8) | | |
| **Pathological stage** | | | | | | 11.106 | 0.004 |
| *I* | 66.7 | 93.3 | 56.0 | 56.0 | 62.5 (40.8-84.1)a | | |
| *II* | 59.3 | 77.8 | 68.8 | 50.0 | 58.6 (44.2-73.0)b | | |
| *III* | 28.6 | 47.6 | 21.8 | 21.8 | 31.2 (13.2-49.2)a,b | | |
| **Histopatologic subtype** | | | | | | 0.645 | 0.724 |
| *Pleomorfik carcinoma* | 49.0 | 71.4 | 49.5 | 38.3 | 49.1 (37.1-61.1) | | |
| *Spindle and/or giant cell ca.* | 50.0 | 62.5 | 46.9 | 46.9 | 45.7 (19.4-72.1) | | |
| *Carcinosarcoma* | 66.7 | 83.3 | 66.7 | 66.7 | 71.2 (36.2-106.1) | | |
| **PD-L1** | | | | | | 2.750 | 0.097 |
| *Negative* | 61.1 | 94.4 | 69.6 | 47.7 | 66.0 (46.3-85.6) | | |
| *Positive* | 46.7 | 62.2 | 43.1 | 38.8 | 45.7 (32.6-58.8) | | |
| **Overall** | 50.8 | 71.4 | 50.8 | 41.5 | 53.6 (41.9-65.2) | - | - |

CI: Confidence interval, a: Stage I vs III (p=0.015), b: Stage II vs III (p=0.004).

**Table 4.** The results of multiple Cox's regression analysis

|  | RR | 95% CI | p-value |
|---|---|---|---|
| **Pleura invasion** |  |  |  |
| *PL0* | 1,000 | - | - |
| *PL1,PL2,PL3* | 1218 | 0.551-2.692 | 0.626 |
| **Pathological stage** |  |  |  |
| *I* | 1.000 | - | - |
| *II* | 1012 | 0.349-2.938 | 0.983 |
| *III* | 3203 | 1.141-8.996 | **0.027** |
| **PD-L1** |  |  |  |
| *Negative* | 1.000 | - | - |
| *Positive* | 2038 | 0.822-5.052 | 0.124 |

RR: Relative risk, CI: Confidence interval.

Figure Legends

Figure 1. PD-L1 positivity in sarcomatoid carcinomas. A. Spindle cell carcinoma B. Giant cell carcinoma C. Pleomorphic carcinoma D. Strong staining with PD-L1 in pleomorphic carcinoma

Figure 2. Kaplan Meier curves of PD-L1 negative and positive sarcomatoid lung carcinomas

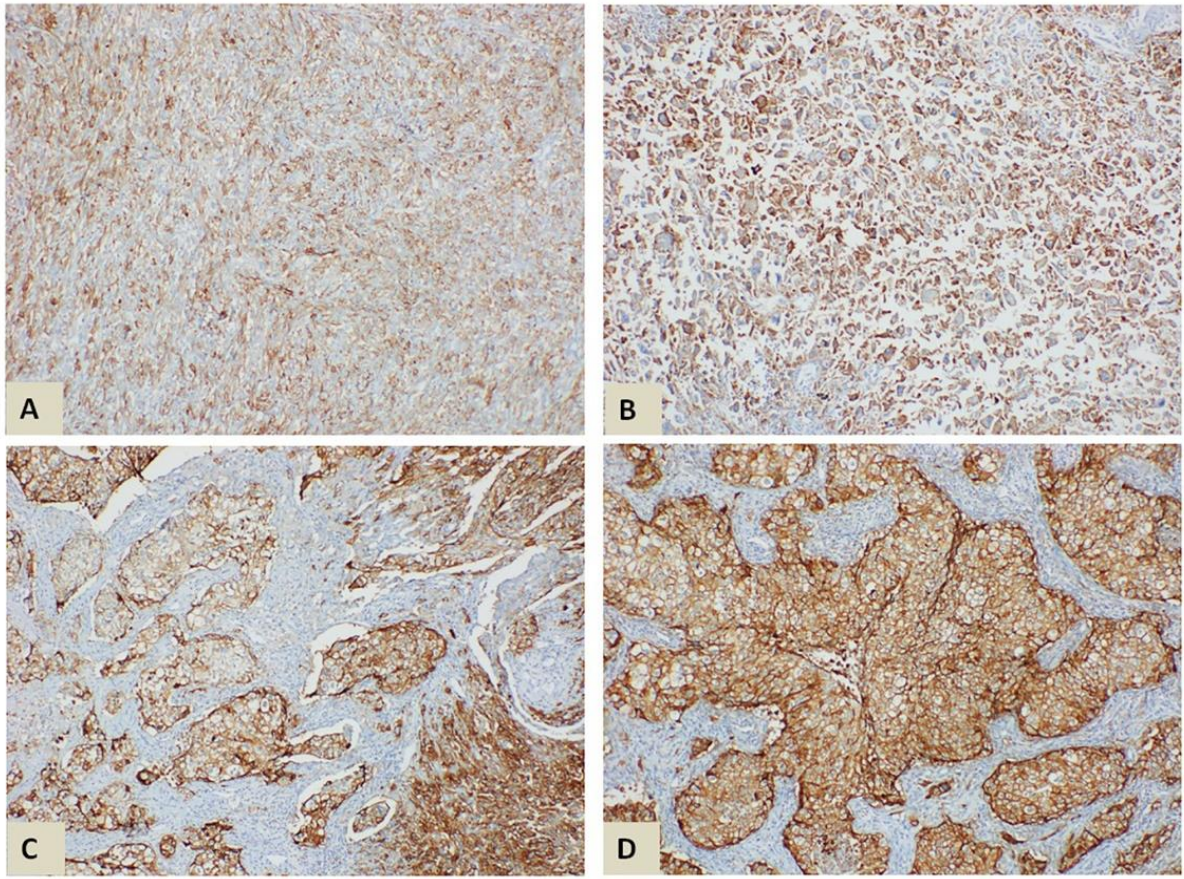

Figure 1. Photomicrographs of PD-L1 (SP263) antibody staining in sarcomatoid lung carcinoma tissue samples. A. Strong staining in spindle-cell carcinoma (100X). B. Strong staining in giant-cell carcinoma (200X). C. Strong staining in pleomorphic carcinoma (200X). D. Higher magnification view of PD-L1-positive pleomorphic carcinoma demonstrating strong uniform staining (400X)

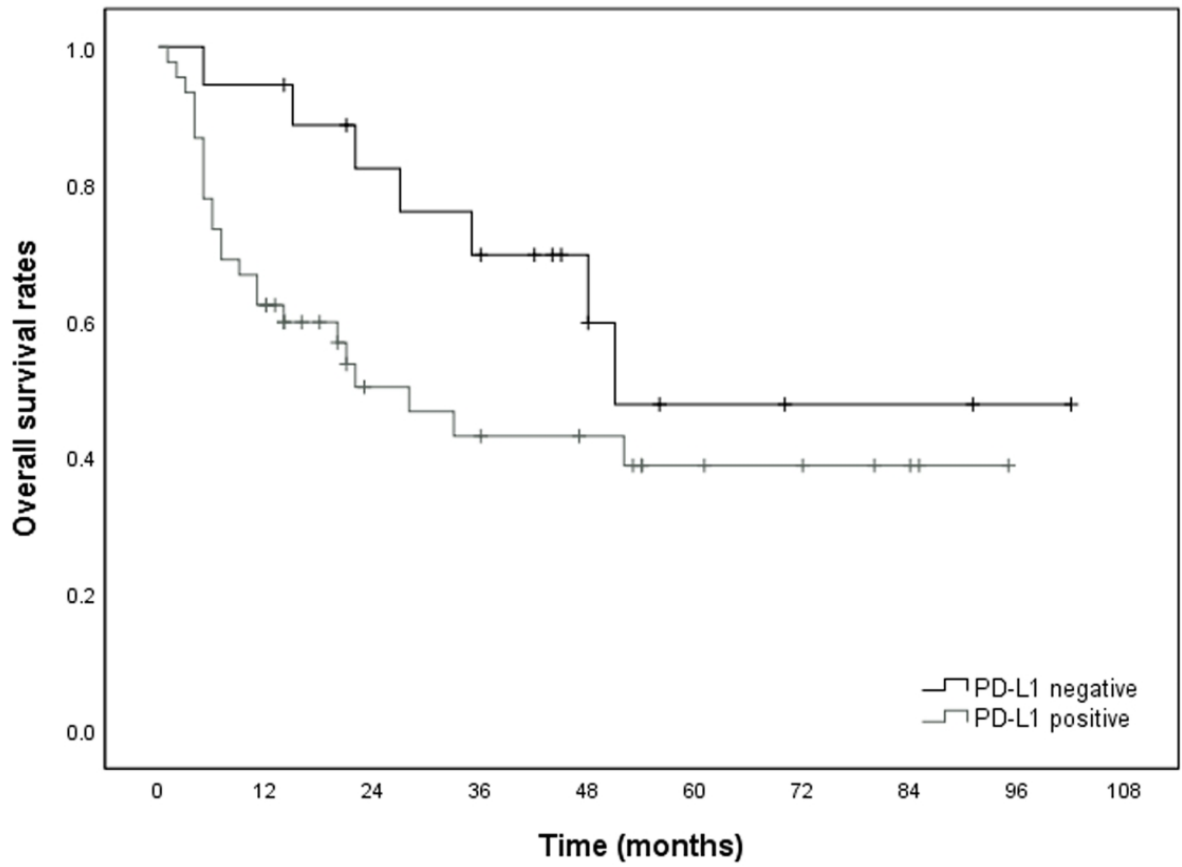

Figure 2. Kaplan Meier curves of PD-L1 negative and positive sarcomatoid lung carcinomas
81x60mm (600 x 600 DPI)